\documentstyle[12pt,epsf]{article}
\begin{document}
\newpage
\baselineskip 16pt plus 2pt minus 2pt

\thispagestyle{empty}

\par
\topmargin=-1cm      

{ \small

\noindent{KFA-IKP(TH)-1996-08} \hfill{LPT 96-20}

\noindent{DOE/ER/40427-30-N96} \hfill{DUKE-TH-96-131}
}

\vspace{30.0pt}

\begin{centering}
{\Large\bf Neutral pion photoproduction on deuterium in
baryon chiral perturbation theory to order $q^4$}\\

\vspace{40.0pt}
{{\bf S.R.~Beane}$^1$,
{\bf V.~Bernard}$^2$,
{\bf T.-S.H.~Lee}$^3$\\
{\bf Ulf-G.~Mei{\ss}ner}$^4$,
{\bf U.~van Kolck}$^5$ }\\
\vspace{20.0pt}
{\sl $^{1}$Department of Physics,
Duke University, Durham, NC 27708, USA} \\ and \\
{\sl Department of Physics, University of Maryland,
College Park, MD 20742, USA}\\
{\it E-mail address: sbeane@fermi.umd.edu}\\  
\vspace{15.0pt}
{\sl $^{2}$Laboratoire de Physique Th\'eorique} \\
{\sl Universit\'e Louis Pasteur, F-67037 Strasbourg Cedex 2, France} \\
{\it E-mail address: bernard@crnhp4.in2p3.fr}\\
\vspace{15.0pt}
{\sl $^{3}$Physics Division,
Argonne National Laboratory, Argonne, IL 60439, USA} \\
{\it E-mail address: lee@phy.anl.gov}\\ 
\vspace{15.0pt}
{\sl $^{4}$Institut f\"ur Kernphysik, Forschungszentrum J\"ulich, 
D-52425 J\"ulich,
Germany} \\
{\it E-mail address: Ulf-G.Meissner@kfa-juelich.de}\\
\vspace{15.0pt}
{\sl $^{5}$Department of Physics,
University of Washington, Seattle, WA 98195-1560, USA} \\
{\it E-mail address: vankolck@phys.washington.edu}\\
\end{centering}
\vspace{20.0pt}
\begin{center}
\begin{abstract}
Threshold neutral pion photoproduction on the deuteron is studied in
the framework of baryon chiral perturbation theory beyond next--to--leading 
order in the chiral expansion. To fourth order in small momenta, the 
amplitude is finite and a sum of two-- and three--body interactions with no
undetermined parameters.  With accurate theoretical and experimental
input from the single nucleon sector for the proton amplitude, we 
investigate the sensitivity of the threshold cross section to the 
elementary $\gamma n \to \pi^0 n$ amplitude. A precise measurement of the
threshold cross section for $\gamma d \to \pi^0 d$ is called for.
\end{abstract}

\vspace*{10pt}

\vfill
\end{center}

\newpage
\section{Introduction}

One of the major goals in nuclear physics is the understanding of
isospin symmetry violation related to the light quark mass
difference $m_u -m_d$ and virtual photon effects. Although the
light quark mass ratio deviates strongly from unity, 
$m_d/m_u \sim 2$ \cite{gl75} and one thus could expect sizeable isospin 
violation, such effects are effectively masked since  $m_d - m_u \ll
\Lambda$, with $\Lambda$ the scale of the strong interactions (which can be
chosen to be $4\pi F_\pi$ or 1~GeV or the mass of the $\rho$). To assess
the isospin violation through quark mass differences,
precise measurements and accurate calculations are mandatory. As pointed out
by Weinberg a long time ago \cite{weinmass}, systems involving nucleons
can exhibit such effects to leading order in contrast to the suppression
in purely pionic  processes due to G--parity. Pion--nucleon scattering or 
neutral pion photoproduction are best suited for such investigations since
a large body of (precise) data exists for various isospin channels. While
there is still some discrepancy about the low--energy $\pi^- p$ elastic 
scattering data, very precise neutral pion photoproduction experiments
have been performed over the last few years at MAMI \cite{fuchs}
and SAL \cite{jack}. Together with the new data on $\gamma p \to \pi^+ n$,
a sensitive test of isospin symmetry could be performed
if one would have information on the elementary neutron amplitude $\gamma
n \to \pi^0 n$. This amplitude can only be inferred indirectly from 
reactions involving few--nucleon systems like the deuteron or $^3$He. In this
paper, we concentrate on coherent neutral pion production off deuterium in 
the threshold region and study the sensitivity of the deuteron electric dipole 
amplitude $E_d$ to the elementary neutron amplitude, $E_{0+}^{\pi^0n}$.
The framework to do this is
baryon chiral perturbation theory, which is reviewed in great detail in
Ref.\cite{bkmrev}. It is the effective field theory of the standard model
at energies below 1~GeV
and allows one to explore in a systematic fashion the strictures of the 
spontaneously and explicitly broken chiral symmetry underlying the 
fundamental color gauge theory of the strong interactions, QCD. While 
originally formulated for the Goldstone boson sector, the machinery
can be extended straightforwardly to include single--baryon processes,
treating the baryons as very heavy, static sources \cite{jm}\cite{bkkm}.
In few--nucleon systems, a complication arises due to the existence of
shallow nuclear bound states \cite{weinnp} and the related   
infrared singularities in Feynman diagrams evaluated in the static 
approximation. One way to overcome this is to  
adapt the rules and use chiral perturbation theory to
calculate an effective potential, which consists of the sum of all
$A$-nucleon irreducible graphs \cite{swnp1}.  The
S-matrix, which of course includes all reducible contributions, is then
obtained through iteration by solving a Lippmann-Schwinger equation. Several
generic features of nuclear physics, in particular the relative
strengths of two- and few-body forces \cite{weinnp}\cite{ubi1}\cite{ubi2}
or the dominance of soft--pion exchange currents \cite{rho},
have been shown to arise naturally in this approach.
For calculating scattering processes involving light nuclei, 
Weinberg \cite{swnp3} proposed to use chiral perturbation theory to
generate the irreducible kernel and combine these with very accurate 
phenomenological
external nuclear wave functions. Although nuclear wave functions
are calculable in chiral perturbation theory \cite{ubi1}, they have
not yet been determined with the accuracy matching that of the
phenomenological models.  
Therefore, we use the input from the successful phenomenological
boson--exchange potentials which differ mostly in their short--range 
interaction parts. Using a large variety of these potentials allows one to
assess to which degree of accuracy one is sensitive to the chiral symmetry
constraints used in determining the irreducible scattering kernel.

In this paper, we consider pion photoproduction on the deuteron at
threshold extending the previous work by Beane et al.\cite{silas}.
We calculate the invariant threshold amplitude to
fourth order in small momenta and/or pion masses. This is mandated by the
fact that only to this order in the chiral expansion one can achieve
a satisfactory description of the elementary process $\gamma p \to \pi^0 p$
\cite{bkmz}\cite{bkmplb}. As already stressed in Ref.\cite{silas},
to third  order in small momenta,
the interaction kernel for neutral pion photoproduction off the deuteron
has no undetermined
parameters. Unfortunately, although the same is true in the single
nucleon sector, evidently the amplitude there converges slowly at
best and thus the single-scattering contribution was treated as
phenomenological input. Since that paper was published, the situation has 
considerably improved and we are now in the position to consistently include
the predictions for the single--nucleon sector. In particular, the precise
${\cal O}(q^4)$ (where $q$ is a generic symbol for a small momentum or meson 
mass) chiral perturbation theory calculation \cite{bkmz}\cite{bkmplb}
predicts a very large neutron electric dipole amplitude, larger in magnitude
than the corresponding proton one. As we will show, to fourth order in small
momenta the interaction kernel can still be given entirely in terms of known 
parameters, and thus we will be able to calculate the sensitivity of the
threshold cross section to the elementary neutron--$\pi^0$ amplitude. We
mention that charged pion photoproduction at threshold has been 
calculated accurately (to four orders) in chiral perturbation 
theory~\cite{bkmkr}.

Since neutral pion photoproduction has already been studied extensively
in more conventional nuclear approaches based e.g. on meson-exchange
models (see e.g. refs.\cite{koch}--\cite{ktb}), 
it is worth emphasizing why we use chiral perturbation theory here.
As first stressed by Weinberg \cite{swnp3}, chiral perturbation theory 
allows one to {\it systematically} construct the many--body interactions 
between nucleons, pions and photons following the power counting rules.
For example, in the case of elastic pion--deuteron scattering the 
phenomenologically dominant rescattering contribution was shown to be the
leading three--body interaction in the chiral expansion thus leading to 
deeper theoretical insight. 
Here, the phenomenological approaches to the single neutral 
pion production amplitude off protons  based on Born terms are at variance 
with the data by many standard deviations quite in contrast to 
chiral perturbation theory. To tackle the problem of how sensitively the
deuteron threshold cross section depends on the elementary neutron--$\pi^0$
amplitude can therefore only be addressed in a framework which allows one
to systematically include and order the various contributions arising from
single and multiple scattering processes. It goes without saying that
this approach is only useful for the threshold region. We remark that 
contributions from the $\Delta$--isobar are encoded in
some of  the low--energy constants appearing in the single scattering
matrix elements.

This paper is organized as follows. In section 2 we briefly review the
effective Lagrangian underlying the calculation and the
standard power counting formulas.  In section 3 the calculation of the
various contributions to the threshold S--wave amplitude is outlined.
Section 4 contains the results and discussions thereof.  
The appendices include our conventions and give some more details 
on the calculations.

\section{Effective field theory}

In this section, we briefly discuss the effective chiral Lagrangian
underlying our calculations and the corresponding power counting. For
more details we refer to the review \cite{bkmrev} and to the paper by 
Beane et al.\cite{silas}.

At low energies, the relevant degrees of freedom are hadrons, in 
particular the Goldstone bosons linked to the spontaneous symmetry
violation. We consider here the two flavor case and thus deal with the
triplet of pions, collected in the matrix $U(x) = \xi^2(x)$.
It is straightforward to build an effective Lagrangian
to describe their interactions, called ${\cal L}_{\pi\pi}$. This Lagrangian
admits a dual expansion in small (external) momenta and quark (meson) 
masses as detailed below. Matter fields such as nucleons can also be 
included in the effective field theory based on the
familiar notions of non--linearly realized chiral symmetry.
These pertinent effective Lagrangian splits into two parts,
${\cal L}_{\pi N}$ and ${\cal L}_{NN}$, with the first (second) one
consisting of terms with  exactly one (two) nucleon(s) in the initial 
and the final state. Terms with more nucleon fields are of no relevance
to our calculation. The pertinent contributions to neutral pion 
photoproduction at threshold are organized according to the standard
power counting rules, which for a generic matrix element involving the
interaction of any number of pions and nucleons can then be written in
the form
\begin{equation}
{\cal M}={q^\nu}{\cal F}(q/\mu),
\end{equation}
where $\mu$ is a renormalization scale, and $\nu$ is a counting index, i.e.
the chiral dimension of any Feynman graph. $\nu$ is, of course, intimately
connected to the chiral dimension $d_i$ which orders the various terms in the
underlying effective Lagrangian (for details, see \cite{bkmrev}). 
For processes with the same number of nucleon lines
in the initial and final state $(A)$, one finds\cite{weinnp}
\begin{eqnarray}
\nu &=&4-{A}-2C+2L+{\sum _i}{V_i}{\Delta _i}\nonumber \\
& &{\Delta _i}\equiv {d_i}+{n_i}/2-2.
\label{index}
\end{eqnarray}
where $L$ is the number of loops,
$V_i$ is the number of vertices of type $i$, $d_i$ is
the number of derivatives or powers of $M_\pi$ which contribute to an
interaction of type $i$ with $n_i$ nucleon fields, and $C$ is the number of
separately connected pieces.
This formula is important because chiral symmetry places a lower
bound: ${\Delta _i}\geq 0$. Hence the leading {\it irreducible} graphs
are tree graphs ($L=0$) with the maximum number $C$ of separately
connected pieces, constructed from vertices with $\Delta _i =0$.
In the presence of an electromagnetic field, this formula is
slightly modified. Photons couple via the electromagnetic
field strength tensor and by minimal substitution.
This has the simple effect of modifying the lower bound on $\Delta _i$ to
${\Delta _i}\geq -1$, and of introducing an expansion in
the electromagnetic coupling, $e$. Throughout, we work to first order in $e$,
with one exception to be discussed below.
In what follows, we will work within the one--loop approximation to order
$q^4$ (notice that we refer here to the chiral dimension used to organize
the various terms in the calculation of the single--nucleon 
photoproduction amplitudes), 
which extends the ${\cal O}(q^3)$ calculation of ref.\cite{silas}.
Such a higher order calculation is mandated by the fact that to order $q^3$
the single--nucleon neutral pion photoproduction amplitudes are too 
inaccurate. Furthermore, one would like to see how big the corrections
to the leading order  three--body interactions calculated in ref.\cite{silas}
are. In terms of the counting index $\nu$, in ref.\cite{silas} all terms
with $\nu = 4-3A=-2$ and   $\nu = 5-3A=-1$ were included. We go one
order further, i.e. we add {\it all} terms with $\nu = 6-3A=0$ (remember that  
$A=2$ and we have exactly one photon coupling with $\Delta_i = -1$).  
Consequently, the effective Lagrangian consists of the following pieces:
\begin{equation}
{\cal L}_{\rm eff} = {\cal L}_{\pi\pi}^{(2)} + {\cal L}_{\pi N}^{(1)}
 + {\cal L}_{\pi N}^{(2)}  + {\cal L}_{\pi N}^{(3)}
 + {\cal L}_{\pi N}^{(4)}  + {\cal L}_{N N}^{(0)}
 + {\cal L}_{N N}^{(1)} \,\, ,
\end{equation}
where the index $(i)$ gives the chiral dimension $d_i$ (number of derivative
and/or meson mass insertions).
The form of ${\cal L}_{\pi\pi}^{(2)}+ {\cal L}_{\pi N}^{(1)} $ is standard.
The terms from  ${\cal L}_{\pi N}^{(3)} + {\cal L}_{\pi N}^{(4)}$ 
contributing to the single--nucleon photoproduction amplitudes 
are given in ref.\cite{bkmz}. Concerning the terms from  
${\cal L}_{\pi N}^{(2)}$ and ${\cal L}_{N N}^{(1)}$, we will discuss
the pertinent ones below.\footnote{Of course, terms from   
${\cal L}_{\pi N}^{(2)}$  also appear in the
single nucleon calculation of refs.\cite{bkmz}\cite{bkmplb} and from
${\cal L}_{NN}^{(1)}$ in the nuclear force calculation of ref.\cite{ubi2}.}
With these at hand, we have to calculate tree and loop graphs
for the single--scattering amplitude and tree graphs with exactly one insertion
from ${\cal L}_{\pi N}^{(2)}$ for the three--body interactions between 
pions, nucleons and photons as well as tree graphs with one insertion from
${\cal L}_{N N}^{(1)}$. As we will show later, the only new coupling constants
appearing at ${\cal O}(q^4)$ are related to the single--nucleon sector and
have already been determined in refs.\cite{bkmz}\cite{bkmplb}. After these 
general remarks, let us now turn to the actual calculations.

\section{Anatomy of the calculation}

In this section, we outline how the various contributions to the
electric dipole amplitude $E_d$ are calculated. The pertinent details
are relegated to the appendices, including our conventions and definitions
of the S--matrix.

\subsection{Threshold cross section}

Consider the reaction $\gamma (k) + d(p_1) \to \pi^0 (q) + d(p_2)$
in the threshold region, $\vec{q} \simeq 0$.  For real photons
($\vec{\epsilon} \cdot \vec{k} = 0$, with $\vec \epsilon$ the photon 
polarization vector) the S--wave photoproduction amplitude can be expressed
in terms of two multipoles as,
\begin{eqnarray}
{\cal M}_d &=& 16 i \pi (m_d +M_\pi) \, \int d^3p \,
{{\phi}}^{\ast}_{f}(\vec{p}) \, \,
 \biggl\{
{\cal M}_1 \, \frac{1}{2} (\vec{\sigma}_1 + \vec{\sigma}_2 )
\cdot \vec{\epsilon} \nonumber \\
&+& {\cal M}_2 \, \frac{1}{2}
 \, [i\, \vec{\sigma}_1 \cdot \hat{k} \, 
\vec{\sigma}_2 \cdot (\vec{\epsilon}\times \hat{k}\,) + (1 \leftrightarrow 2
)] \biggr\} \, {{\phi}}_{i}(\vec{p}-\vec{k}/2)
\,\, .
\label{M12}
\end{eqnarray}
The differential cross section at threshold takes the form (see appendix~A)
\begin{equation}
\frac{|{\vec k}|}{|{\vec q\,}|}\;\frac{d\sigma}{d\Omega}
\bigg|_{|{\vec q\,}|=0} =\frac{8}{3}{E_d^2}
\, \, .
\end{equation}
In what follows, we will discuss the chiral expansion of the dipole
amplitude $E_d$. In appendix~A we discuss the relation to the commonly
used amplitudes treating the deuteron as an elementary particle.

\subsection{Single scattering contribution}

The single scattering contribution is given by all diagrams where the
photon and the pion are absorbed and emitted, respectively, from one
nucleon with the second nucleon acting as a spectator (the so--called
impulse approximation). One only has a 
contribution to the ${\cal M}_1$ amplitude of the form 
\begin{eqnarray} \label{Edss}
{E^{ss}_{d}} &=&
\frac{1+{M_\pi}/{m}} {1+{M_\pi}/{m_d}} \, \biggl\{ \frac{1}{2} \,
({E_{0+}^{{\pi ^0}p}}+{E_{0+}^{{\pi ^0}n}} ) \, \int d^{3}{p} \; 
{{\phi}}^{\ast}_{f}(\vec{p}) \, \vec{\epsilon} \cdot \vec{J} \,
{{\phi}}_{i}(\vec{p}-\vec{k}/2) \nonumber \\
&-& \frac{k}{m} \, \hat{k} \cdot \int d^{3}{p} \; \hat{p} \,
\frac{1}{2} \, ({P_{1}^{{\pi ^0}p}}+{P_{1}^{{\pi ^0}n}} ) \,
{{\phi}}^{\ast}_{f}(\vec{p})\, \vec{\epsilon} \cdot \vec{J} \,
{{\phi}}_{i}(\vec{p}-\vec{k}/2) \biggr\} \,\, ,
\end{eqnarray}
evaluated at the threshold value
\begin{equation}
|\vec{k}| =  k_{\rm thr} 
= M_{\pi^0} - \frac{M_{\pi^0}^2}{2\,m_d} = 130.1 \,{\rm MeV} \,\, ,
\label{valk}
\end{equation}
and with $\vec J =  (\vec{\sigma}_1 + \vec{\sigma}_2 )/2$.
A number of remarks concerning  Eq.(\ref{Edss}) are in order. It is important
to differentiate between the $\pi^0 d$ and the $\pi^0 N$ ($N=p,n$) 
center--of--mass (COM) systems. 
At threshold in the former, the pion is produced at 
rest, it has, however, a small three--momentum in the latter \cite{koch}.
Consequently, one has a single--nucleon P--wave contribution proportional
to the elementary amplitudes $P_{1}^{{\pi ^0}p}$ and $P_{1}^{{\pi ^0}n}$
as defined in \cite{bkmz}. We use the P--wave low--energy theorems
found in that paper,
\begin{equation}\label{Plets}
P_{1}^{{\pi ^0}p} = 0.480 \, |\vec{q}\,| \,{\rm GeV}^{-2} \, \quad
P_{1}^{{\pi ^0}n} = 0.344 \, |\vec{q}\,| \,{\rm GeV}^{-2} \, \, ,
\end{equation}
with
\begin{equation}\label{boost}      
\vec{q} = \mu \, (1 - \mu ) \, \vec{p} -  \mu^2 \,
m \, (1 - 5\mu / 4 ) \, \hat{k} /2 \,\,\, ,
\end{equation}
with $\mu = M_\pi / m$ and $\vec{p}$ is the nucleon three--momentum
in the $\pi^0 d$ COM system. The derivation of Eq.(\ref{boost}) is sketched
in appendix~B. We have checked that dropping the terms of order
$M_\pi^2$ in Eq.(\ref{boost}) does not alter the results within the
precision given.  Furthermore, we neglect the
energy dependence of the elementary $\pi^0 p$ and $\pi^0 n$ 
S-- and P--wave amplitudes since the pion energy changes only 
by 0.4\% for typical
average nucleon momenta in the deuteron (see appendix~B).
We take for  the elementary S--wave pion production 
amplitudes the predictions from the ${\cal O}(q^4)$ chiral 
perturbation theory calculation \cite{bkmplb},
\begin{equation}
E_{0+}^{{\pi ^0}p} = -1.16 \times 10^{-3}/M_{\pi^+} \, , \quad
E_{0+}^{{\pi ^0}n} = +2.13 \times 10^{-3}/M_{\pi^+} \, .
\end{equation}
In that calculation, the dominant isospin breaking effect, namely the
charged to neutral pion mass difference, which is 
almost entirely of electromagnetic
origin, has been taken into account. A fully consistent calculation 
including all effects from virtual photons and the quark mass differences is
not yet available. The result for $E_{0+}^{{\pi ^0}n}$ is based on the
assumption that the counter terms entering at order $q^4$ are the same
for the proton and the neutron apart from trivial isospin factors (for a 
detailed discussion, see ref.\cite{bkmz}).
We also note that only the first process has been measured, the recent 
determinations from MAMI and SAL give
\begin{equation}
E_{0+}^{{\pi ^0}p} = (-1.31 \pm 0.08) \times 10^{-3}/M_{\pi^+} 
\cite{fuchs}\, , \;\;\;
E_{0+}^{{\pi ^0}p} = (-1.32 \pm 0.08) \times 10^{-3}/M_{\pi^+} 
\cite{jack} \, .
\end{equation}
Furthermore, the P--wave LET for neutral pion production off protons has
been shown to hold within 3\% in Ref.\cite{fuchs}.
Using the Argonne V18 \cite{v18}, the Reid soft core (RSC) \cite{reid}, the
Nijmegen \cite{nij} and the Paris \cite{paris} potential, we evaluate
$E_d^{ss}$ and find
\begin{equation} \label{Edssval}
{E^{ss}_{d}}=
{0.36\times {10^{-3}}}/{{{M_{\pi^+}}}},
\end{equation}
with an uncertainty of $\delta {E^{ss}_{d}} = {0.05\times {10^{-3}}}/
{{{M_{\pi^+}}}}$ due to the various potentials used. The P--wave contribution
amounts to a 3\% correction to the one from the S--wave, i.e. it amounts to
a minor correction. The sensitivity of the single--scattering contribution 
$E_d^{ss}$ to the elementary neutron--$\pi^0$ amplitude is given by
\begin{equation}
E_d^{ss} = 0.36 - 0.38 \cdot (2.13 - E_{0+}^{{\pi ^0}n}) \; , 
\label{sensi}
\end{equation}
all in units of $10^{-3}/M_{\pi^+}$. Consequently, for $E_{0+}^{{\pi ^0}n}
= 0$, we have $E_d^{ss} = -0.45$ which is of opposite sign to the value
based on the chiral perturbation theory prediction for $E_{0+}^{{\pi ^0}n}$. 
If one were to use the empirical value for the proton
amplitude, the single--scattering contribution would be somewhat reduced.

Finally, we remark that the single--scattering contribution given
in ref.\cite{silas} is very different. This is due to the following.
First,  in that paper a factorized form for the single--scattering
contribution was used,
\begin{equation}\label{Edssfac}
{E^{ss}_{d}}=
\frac{1+{M_\pi}/{m}} {1+{M_\pi}/{m_d}} \, \frac{1}{2} \,
({E_{0+}^{{\pi ^0}p}}+{E_{0+}^{{\pi ^0}n}} ) \, {{S_d}(k_{\rm thr}/2)}=
{0.41\times {10^{-3}}}/{{{M_{\pi^+}}}},
\end{equation}
where ${S_d}(k_{\rm thr}/2)$ is the deuteron form-factor,
\begin{equation}
{S_d}(k_{\rm thr}/2)=\int d^{3}{p} \; {{\phi}}^{\ast}_{f}(\vec{p})
\;{{\phi}}_{i}(\vec{p}-\vec{k}_{\rm thr}/2)=0.79 \, \, .
\label{dff}
\end{equation}
Note that the isospin factor $1/2$  was inadvertently omitted in \cite{silas}.
Such a form is strictly correct only for the S--wave part of the deuteron 
wave function. Switching off the D--wave component of the deuteron wave 
function, we get the same result as in \cite{silas}. The main reason for the
large numerical difference of our single--scattering contribution to theirs  
can be traced back to the
use of the then accepted empirical value of $E_{0+}^{{\pi ^0}p} 
= (-2.1 \pm 0.2) \times 10^{-3}/M_{\pi^+}$ and the one from the 
incomplete low--energy theorem for $\gamma n \to \pi^0 n$,
$E_{0+}^{{\pi ^0}n} \simeq 0.5 \times 10^{-3}/M_{\pi^+}$ in \cite{silas} .

\subsection{Three--body contributions at order $q^3$ ($\nu \le -1$)}

Although the ${\cal O}(q^3)$ (corresponding to the counting index
$\nu =-2$ and $-1$) three--body contributions (meson exchange currents)
have been worked out in ref.\cite{silas}, we will give them here for 
completeness. In the
Coulomb gauge ($\epsilon \cdot v=0$, with $v_\mu$ the nucleons' 
four--velocity) and at threshold, only the two diagrams shown in Fig.~1
contribute. In momentum space, the corresponding structures are
\begin{equation}
a: \,\, \frac{1}{\vec{q} \,^2} \, , \quad
b: \,\, \frac{(q-k)_i \, q_j}{[(\vec{q}  -\vec{k}\,)^2 +
M_\pi^2] \, \vec{q} \,^2} \,\,\,\, ,
\end{equation}
where $\vec{q} = \vec{p} - \vec{p\,}'$ 
with $\vec{p}$ and $\vec{p}\,'$ the nucleon cms three--momenta in the
initial and the final state, respectively. These expressions can be 
Fourier transformed into coordinate space easily and one finds
that they give a contribution to the ${\cal M}_1$ amplitude defined
in Eq.(\ref{M12}) \cite{silas}.
Evaluating these with standard deuteron wavefunctions gives,
\begin{equation}
E_d^{(i)} = E_d^{(i,S)} + E_d^{(i,SD)} + E_d^{(i,D)} \quad (i=a,b)
\end{equation}
where the indices $S,SD,D$ refer to  the contribution from
the S-wave, the mixed and the D--wave part of the deuteron wavefunction.
We find for the three--body (tb) contribution at order $q^3$
\begin{equation}
E_d^{tb,3} = (-1.90\, ,\, -1.88 \, , \, -1.85 \, , \, -1.88)
\times 10^{-3}/M_{\pi^+} \,\, ,
\end{equation}
for the Argonne V18, RSC, super soft core (SSC) \cite{ssc} and 
Bonn potential \cite{bonn} \cite{mach}, in order, 
and using $g_A = 1.32$ as determined from the Goldberger--Treiman relation
(for $g_{\pi N} =13.4$) (to be consistent with the calculation of the
elementary amplitudes \cite{bkmplb}). Note that
we use the full (energy-dependent) model of ref.\cite{bonn} throughout. 
The number for the
Bonn potential differs from the one given in ref.\cite{silas} for three
reasons. First, the overall sign of the contribution from graph b was
incorrectly given and second, we find that the SD-- and D--wave contributions
from this graph substantially alter the pure S--wave contribution. Also,
the D--wave contributions to graphs a and b were incorrectly evaluated
(the corrected formulae are given in appendix~C). If we,
however, only retain the S--wave part of the deuteron wavefunction, we
recover the result of Eq.(26) of ref.\cite{silas} (modulo the sign and within
the numerical precision).

\subsection{Three--body contributions at order $q^4$ ($\nu=0$)}

At this order, we have to consider tree graphs with exactly one
insertion from the (chiral) dimension two pion--nucleon 
Lagrangian, which has the
general form (in the isospin limit $m_u =m_d$)
\begin{eqnarray}
\label{LpiN2}
{\cal L}_{\pi N}^{(2)}&=&{\bar {H}}\biggl\{-\frac{1}{2m}{D^2}
+ \frac{1}{2m}  (v\cdot D)^2
+\frac{i\,g_A}{2m}\{{v\cdot A},{S\cdot D}\} \nonumber \\
& &-\frac{i}{4m}[{S_v^\mu},{S_v^\nu}]\,[(1+{\kappa _v}){f_{\mu\nu}^+}+
\frac{1}{2}({\kappa _s}-{\kappa _v}){\rm Tr}({f_{\mu\nu}^+})] \\
& &+ c_1 \,{\rm Tr}(\chi_+) + \biggl(c_2-\frac{g_A^2}{8m}\biggr)(v \cdot u)^2
+ c_3 \, u \cdot u + \biggl(c_4+\frac{1}{4m}\biggr)[S^\mu,S^\nu]u_\mu u_\nu
\, \biggr\} {H}, \nonumber
\end{eqnarray}
where $m$ is the nucleon mass and  
${f_{\mu\nu}^+}\equiv e({\xi ^\dagger}Q\xi + \xi Q{\xi ^\dagger})
{F_{\mu\nu}}$.  ${F_{\mu\nu}}$ is the electromagnetic
field strength tensor and $S^\mu$ is the covariant spin--operator. $H$
denotes the large component of the nucleon spinor (for more details, see 
ref.\cite{bkmrev}). The terms in ${\cal L}_{\pi N}^{(2)}$ fall into two 
classes, the first one with fixed coefficients due to 
Lorentz invariance\cite{bkkm} and the second one with low--energy constants
$c_1 , \ldots,c_4$ and two other constants that are directly related to the
isoscalar and isovector nucleon anomalous magnetic moments,
${\kappa _v}={\kappa _p}-{\kappa _n}$, ${\kappa _s}={\kappa _p}+
{\kappa _n}$. The pertinent 
three--body interaction diagrams are shown in Fig.~2, the blob characterizing
an insertion from ${\cal L}_{\pi N}^{(2)}$. In principle, all the terms
appearing in Eq.(\ref{LpiN2}) can contribute. In addition, there are the
recoil corrections to the graphs in Fig.~1.

Consider first insertions proportional to the low--energy constants
$c_{1,2,3,4}$. One can show that all of corresponding terms vanish
at threshold either due to the selection rule $S \cdot q=0$ or due to
some isospin factor of the type $\delta^{a3}\,\epsilon^{a3b}$. We are thus 
left with insertions from the terms $\sim 1/2m$, $\sim g_A/2m$ and $\sim 
\kappa_{v,s}$. All of these can be classified in momentum space according
to the following structures,
\begin{eqnarray}
& &\frac{1}{\vec{q\,}^2} \, , \quad \frac{q_i}{\vec{q\,}^2} \, , \quad 
\frac{q_i\, q_j}{\vec{q\,}^2} \, , \quad 
\frac{q_i\, (p+p')_j}{\vec{q\,}^2} \, , 
\frac{(q-k)_i \, (p+p')_j }{[(\vec{q} -\vec{k}\,)^2 + M_\pi^2] } \, , \quad 
\nonumber \\
& &\frac{(q-k)_i \, q_j }{[(\vec{q} -\vec{k}\,)^2 + M_\pi^2] } \, , \quad 
\frac{(q-k)_i \, q_j \, 
({\vec{p}\,}^2 - {\vec{p}\,'}^2)}{[(\vec{q} -\vec{k}\,)^2 +
M_\pi^2] \, \vec{q\,}^2} \, , \quad 
\frac{(q-k)_i \,}{[(\vec{q} -\vec{k}\,)^2 + M_\pi^2]}  \, , \quad
\nonumber \\ 
& &\frac{(q-k)_i \,q_j}{[(\vec{q} -\vec{k}\,)^2 + M_\pi^2]^{3/2}}  \, , \quad 
\frac{(q-k)_i \,(q-k)_j}{[(\vec{q} -\vec{k}\,)^2 + M_\pi^2]^{3/2}} \,\,\, .
\end{eqnarray}
All these operators can 
straightforwardly be transformed into coordinate space and the angular
integrations can be performed analytically leading to the form given
in Eq.(\ref{M12}) and one is left with one simple radial integration. Only
the second operator in the second line is more easily evaluated in 
momentum space. Whenever possible, we have performed both types of 
integration as a check on our numerics (for details, see appendix~C). 
In fact, there is a technical problem with the contribution from graphs b,
l and p. Counting powers of momenta, one sees that they are divergent.
This is due to the fact that in the chiral expansion one has truncated
the pion--nucleon form factor and thus does not suppress the high--momentum
components. This phenomenon also occurs in the calculation of the NN potential
in chiral perturbation theory \cite{ubi1}. As it was done there, we introduce
an additional Gaussian cut--off factor of the form
\begin{equation}
F({\vec q}\,^2) = \exp\{-{\vec q}\,^2 / \Lambda^2_\pi \} \,\,\, ,
\end{equation}
where the cut--off $\Lambda_\pi$ varies between the mass of the $\rho$ and
$4 \pi F_\pi =1.2\,$GeV (since it can be related to the mass scale
of the heavy particles that are integrated out from the effective theory).
Notice that the divergencies appearing in other graphs all
cancel and thus no further regularization is needed for those.
For energy--independent NN potentials as used here, the contribution
from the graphs $n+r$ should be omitted for the reasons explained in detail
in ref.\cite{silas}. We will, however, also calculate their contribution
to get a very rough estimate of
some of the uncertainties due to higher order graphs.
Furthermore, we remark that the kinematical $1/m$ corrections to graph 1a
vanish and the ones for graph 1b are taken care of by shifting the value of
$k_{\rm thr}$ as given in Eq.(\ref{valk}). Finally, appendix~D contains some
details on calculating the time--ordered diagrams in the heavy
baryon approach. 

We find for the three--body (tb) contribution at order $q^4$
\begin{equation}
E_d^{tb,4} = (-0.25\, ,\, -0.23 \, , \, -0.27 )
\times 10^{-3}/M_{\pi^+} \,\, ,
\label{Edtb4}
\end{equation}
for the V18, Reid and SSC potentials, in order, and setting 
$\Lambda_\pi=1\,$GeV. These amount to corrections
of the order of 15\% to the $q^3$ three--body terms.
If one varies the cut--off $\Lambda_\pi$ from $0.65$ to $1.5\,$GeV, 
the three--body
contribution using the V18 potential varies between $-0.24$ and $-0.29$ (in
canonical units), which is a modest cut--off dependence.
The result Eq.(\ref{Edtb4}) is comforting since it shows that
the chiral expansion of the three--body contributions is well under control.
Finally, we remark that if one were to include the contribution from
the recoil graphs $n+r$, the numbers given in Eq.(\ref{Edtb4}) would
change by less than 5 per mille.

\subsection{Contribution from four--nucleon operators $(\nu=0$)}

In the ${\cal O}(q^3)$ calculation of ref.\cite{silas}, no four--nucleon
operators contributed to the deuteron electric dipole amplitude. Using 
Eq.(\ref{index}), we see that for $\nu=0$ we can have exactly one insertion
from ${\cal L}_{NN}^{(1)}$. In Fig.~3a,b, we show the two graphs which in
principle can contribute to pion photoproduction at threshold. 
Clearly, diagram 3a
is only relevant for charged pion photoproduction. Graph 3b also vanishes
for neutral pions, as the following argument shows. This diagram stems from
the one shown in Fig.~3c by applying minimal substitution. The latter
one contains the pion covariant derivative, $\vec{\nabla}_\mu = 
\partial_\mu \vec{\pi} / F_\pi + \ldots$ which upon minimal substitution
takes the form
\begin{equation}
\partial_\mu \pi_a \to \partial_\mu \pi_a - i\,e\,{\cal A}_\mu \, Q_{ab}
\, \pi_b \,\,\, ,
\end{equation}
with 
\begin{equation}
Q_{ab} = i\,\epsilon_{ab3} \,\,\, ,
\end{equation}
the pion charge matrix and ${\cal A}_\mu$ the photon field.
Consequently, $\partial_\mu \pi_3 \to
\partial_\mu \pi_3$ and this type of term can only contribute to charged pion
photoproduction. We therefore conclude that for threshold neutral pion
photoproduction to order $q^4$ (or up to counting index $\nu = 0$) there
is {\it no} contribution from any four--nucleon operator and thus no new, 
a priori undetermined coupling constants appear. In the case of charged pion
photoproduction, this would be different.

\section{Results and discussion}

Since as we have shown in the previous sections, neither the single
scattering nor the three--body corrections depend on the potential
chosen, we will present here results based on the Argonne V18 potential.
The chiral expansion of the electric dipole amplitude $E_d$ takes the 
form
\begin{eqnarray}
E_d &=& E_d^{ss} + E_d^{tb,3} +  E_d^{tb,4}\nonumber \\ 
&=& (0.36  - 1.90 - 0.25) \times 10^{-3}/M_{\pi^+}  
= (-1.8 \pm 0.2)  \times 10^{-3}/M_{\pi^+} \,\,\, .
\label{Edtot}
\end{eqnarray}
It is difficult to estimate the theoretical uncertainty, the value given
in Eq.(\ref{Edtot}) being an educated guess, obtained as follows. We allow 
$E_{0+}^{\pi^0 p}$ to vary between $-1.$ 
and $-1.5$, so that for a fixed value of $E_{0+}^{\pi^0 n} = 2.13$, 
$E_d^{ss}$ varies between 0.2 and 0.4 (all numbers in canonical units).
The results for the three--body contributions are stable at order $q^3$, 
where as we assign a conservative uncertainty of $\pm 0.1$ to the 
corrections at next order due to the cut--off dependence. 
Clearly, this estimate does not include contributions from higher order 
effects, which might give rise to a larger uncertainty.
Our final value is considerably smaller than the  $q^3$ estimate
reported in \cite{silas} for the reasons given above. To see the
sensitivity to the elementary neutron--$\pi^0$ amplitude, we set the
latter to zero and find $E_d = -2.6\times 10^{-3}/M_{\pi^+}$ (V18
potential) which is considerably different from the chiral perturbation
theory prediction, Eq.(\ref{Edtot}), i.e. the S--wave cross section
would differ by a factor of two.
For other values of $E_{0+}^{\pi^0 n}$, $E_d$ can be calculated from
Eq.(\ref{sensi}). Obviously, the sensitivity to the neutron amplitude 
is sizeable and is not completely masked by the larger charge--exchange 
amplitude as it is often stated.

On the experimental side, 
neutral pion photoproduction off deuterium was studied by a group at
Saclay \cite{sacold} and later reanalyzed  in ref.\cite{sac}.
In these papers, the S--wave amplitude ${\cal E}_2$ was extracted.
We remark that this analysis relies heavily on the input 
from the elementary $\pi^0 p$ amplitude (to determine an unknown
normalization factor) and therefore should only be considered indicative. The 
S-- and P--wave multipoles used in \cite{sac} are roughly consistent
with the new determinations from Mainz and Saskatoon.  
The amplitude ${\cal E}_2$ is related to our $E_d$ via
\begin{equation}
|E_d|^2 = |{\cal E}_2|^2 \, \frac{1}{4} \, S_d^2 \, \biggl(
\frac{1+M_\pi /m}{1+M_\pi /2m}\biggr)^2 \,\,\, .
\end{equation}
Using $S_d = 0.79$ and the value of ${\cal E}_2 = (-4.1 \pm 0.4)
\times 10^{-3}/M_{\pi^+}$ \cite{sac}, we have as the ``empirical'' value
\begin{equation}
E_d^{\rm exp} = (-1.7 \pm 0.2) \times 10^{-3}/M_{\pi^+} \,\,\, ,
\label{Edexp}
\end{equation}
taking the same sign as given for ${\cal E}_2$. Since we could not
trace back what the exact value of $S_d$ used in the Saclay analysis
was, we took the value based on the modern potentials evaluated above.
Notice that in the older
papers of the Saclay group, a larger value of  ${\cal E}_2$ was obtained,
based on different input for the proton amplitude (which is at variance
with the new data from Mainz and Saskatoon). This was used to deduce
the empirical number quoted in \cite{silas}. Obviously, this experiment
should be redone at a 100\% duty cycle tagged photon facility, with the
same accuracy as  was done for the process $\gamma p \to \pi^0 p$ at MAMI
and SAL and with a refined theoretical analysis as is available now.
The empirical number Eq.(\ref{Edexp}) agrees nicely with the
theoretical one, Eq.(\ref{Edtot}). However, we remind the reader about
all the assumptions going into the extraction of this ``experimental value''.
Clearly, only a more precise experimental determination of $E_d$ can
tell whether this agreement is of significance. 
On the other hand, if one would find a discrepancy, one would either
have to reassess the calculation of the elementary amplitudes by including
dynamical isospin--breaking effects and/or study in more detail the
wave function dependence of the order $q^4$ three--body corrections.
Finally, two more remarks are in order. First, the sensitivity to the 
neutron amplitude has recently been studied in more conventional 
approaches \cite{lps}\cite{ktb2}. Second, an experiment has been approved
at the Mainz Microtron \cite{mamiexp} to measure the threshold cross
section for coherent neutral pion electroproduction off deuterium at a
photon virtuality of $k^2 = -0.075\,$GeV$^2$.

\bigskip \bigskip

\section*{Acknowledgements}
VB and UGM are grateful to the Nuclear Theory Group at 
Argonne National Laboratory
for hospitality while part of this work was completed.
This research was supported in part by the U. S. Department of Energy,
Nuclear Physics Division
(grants DE-FG05-90ER40592 (SRB), DE-FG02-93ER-40762 (SRB),
 W-31-109-ENG-38 (TSHL) 
and DE-FG06-88ER40427 (UvK)), by  NATO Collaborative Research Grant
950607 (VB, TSHL, UGM) and by the Deutsche Forschungsgemeinschaft
(grant ME 864/11-1 (UGM)).

\newpage

\appendix
\def\theequation{\Alph{section}.\arabic{equation}}
\setcounter{equation}{0}
\section{Invariant amplitudes for the deuteron}

If one considers the deuteron as an elementary particle, the
S--wave photoproduction amplitude can be written as
\begin{equation}
{\cal M}_d = 8 \pi (m_d +M_\pi) \, 2i \,  \vec \epsilon \cdot \vec J \,
\, E_d + {\cal O}(q)\,\,\, ,
\end{equation}
with $m_d$ the deuteron mass and $\vec J = \vec{L} + \vec{S}$ the deuteron 
{\it total} angular momentum,  not simply the sum of the
proton and the neutron spin operators. This is the form commonly
used in analysing the data. Following the conventions used in \cite{silas},
the slope of the differential cross section at threshold takes the form
\begin{eqnarray}
\frac{|{\vec k}|}{|{\vec q}|}\;\frac{d\sigma}{d\Omega}\biggl|_{|{\vec q}|=0}
& = & \frac{1}{64 \pi^2}\;\frac{{\overline{|{\cal M}_d|^2}}}
{(\sqrt{m_{d}^{2}+
M_{\pi^0}^{2}}+|\vec{k}|)\;(m_{d}+M_{\pi^0})}\nonumber\\
&\simeq  & \frac{1}{64{\pi }^{2}}\;
\frac{{\overline{|{\cal M}_d|^2}}}{(m_{d}+M_{\pi^0})^2}\,\,.
\end{eqnarray}
Summing over the final and averaging over the initial states leads to
\begin{eqnarray}
\frac{|{\vec k}|}{|{\vec q}|}\;\frac{d\sigma}{d\Omega}\biggl|_{|{\vec q}|=0}
& = & \frac{2}{2J+1} \sum_{M,M',\lambda} | <JM' |{\vec \epsilon}_\lambda
\cdot {\vec J} \,| JM> \, E_d\,|^2 \nonumber \\
& = & \frac{2}{2J+1} \, \frac{4}{3} \, (2J+1) \, |E_d|^2 
= \frac{8}{3}\, |E_d|^2
\end{eqnarray}
where $M,M'$ are magnetic quantum numbers and $\lambda$ the helicity index
of the photon. We have made use of the Wigner--Eckhardt theorem and 
the fact that $<J||{\vec J}\,||J> = \sqrt{2}$ for the deuteron.


\setcounter{equation}{0}
\section{Two-body to three-body center-of-mass}
\label{cmss}

In this appendix we sketch the derivation of the transformation from
the $\gamma$-$d$ center-of-mass (COM) system to the $\gamma$-$N$ COM.
We are interested in the kinematics of the process
$\gamma{N_1}{N_2}\rightarrow\pi{N_1}{N_2}$, where the nucleons, $N_1$
and $N_2$, are sewn to the deuteron wavefunctions.  Our 3-body
corrections are evaluated in the $\gamma$-$d$ COM whereas the single
scattering corrections which take into account the scattering of the
photon on the individual nucleons have been calculated in the 
$\gamma$-$N$ COM. It is therefore necessary to construct the Lorentz
transformation which boosts the single-scattering corrections to the
$\gamma$-$d$ COM.

We label energy-momenta as $p_1$, $p_2$ and $k$ for $N_1$, $N_2$ and
$\gamma$ in the initial state, respectively, and $p_1'$, $p_2'$ and
$q$ for $N_1$, $N_2$ and $\pi$ in the final state, respectively. The
energy-momentum of the deuteron is given by ${p_\psi}={p_1}+{p_2}$ and
${p_\psi'}={p_1'}+{p_2'}$ in the initial and final state,
respectively. The Fermi momentum, $\vec p$, can then be defined by
${{\vec p}_1}={{\vec p}_\psi}/2+{\vec p}$ and ${{\vec p}_2}={{\vec
p}_\psi}/2-{\vec p}$. Note that with this labelling the initial
deuteron wavefunction has the argument $({{\vec p}_1}-{{\vec
p}_2})/2$=$\vec p$.

The $\gamma$-$d$ COM is defined by ${{\vec p}_\psi}+{\vec k}=0$ and
the $\gamma$-$N_2$ COM is defined by ${{\vec p}_2}+{\vec k}=0$.  The
velocity of the $\gamma$-$N_2$ system in the $\gamma$-$d$ COM is
\begin{equation}
{\vec\beta}=\frac{{{\vec p}_2}+{\vec k}}{E_{2\gamma}}
\end{equation}
where ${E_{2\gamma}}={E_2}+k$.  An arbitrary four-vector, $(E,{\vec
p})$, in the $\gamma$-$d$ COM is expressed in the $\gamma$-$N_2$ COM by
the matrix equation
\begin{equation}
\left(\begin{array}{cc}  {E^*} \\  {p_\parallel^*} \end{array}\right)=
\left(\begin{array}{ccc}  \gamma & {-\beta\gamma} \\ 
               {-\beta\gamma} & \gamma \end{array}\right)
\left(\begin{array}{ccccc}  {E} \\  {p_\parallel} \end{array}\right).
\end{equation}
The $*$-superscript indicates the $\gamma$-$N_2$ COM and ${\vec
p}={{\vec p}_\bot}+{p_\parallel}{\hat\beta}$. We also have ${{\vec
p}_\bot^{\, *}}={{\vec p}_\bot}$.  We can now use this transformation
to find $k$ and $q$ in both frames.  In the text we use a convention
in which the initial deuteron wavefunction has argument ${\vec p}-
{\vec k}/2$. Here this is achieved by the substitution ${\vec
p}\rightarrow -{\vec p}+{\vec k}/2$, such that ${{\vec p}_2}={\vec p}-
{\vec k}$ and ${{\vec p}_1}=-{\vec p}$, since $N_2$ must carry all the
photon momentum.  At the threshold point we find
\begin{eqnarray}
&&{k^*_0}=k_0 - {\vec k}\cdot{\vec p}/m \,\, ,\qquad \quad
{{\vec k}^*}={\vec k} - (k_0/m) \,{\vec p} \,\, ,\nonumber   \\
&&{E^*_\pi}={E_\pi} \, \bigl[1 - {{\vec p \,}^2}/ (2m^2)\bigr] \,\, , \qquad
{{\vec q}^{\, *}}=-E_\pi \, {\vec p} / m  \,\, ,
\end{eqnarray}
where we have included the first non-vanishing $1/m$ corrections. Note
that the pion energy is corrected only at order $1/{m^2}$. Calculating these
corrections with an average nucleon momentum in the deuteron obtained
by means of the uncertainty principle, $\langle \vec p \, \rangle =
46\,$MeV, leads to the 0.4~\% shift mentioned in the main text. Of
course, such an crude estimate does not properly account for binding energy
effects and should be sharpened eventually.

In the $\gamma$-$N_2$ COM we have the multipole decomposition
\begin{equation}
\frac{m}{4 \pi \sqrt{s} } \, T \cdot  \epsilon = 
i \vec \sigma \cdot \vec \epsilon \, (E_{0+}
+ {{\hat k}^*} \cdot {{\hat q}^*} P_1)
+i \vec\sigma \cdot {{\hat k}^*} \, \vec \epsilon \cdot {{\hat q}^*} \,P_2
+ ({{\hat q}^*} \times {{\hat k}^*} ) \cdot \vec \epsilon \, P_3 .
\end{equation}
In terms of the transformed variables it is straightforward to find
that the S--wave multipole is modified by
\begin{equation}
E_{0+}\rightarrow {E_{0+}} - \frac{k}{m}\,{\hat k}\cdot{\hat p}\,{P_1}
\end{equation}
where we have left out terms that vanish upon integration over the
Fermi momentum, and it is understood that the argument of $P_1$ is
a transformed pion momentum.

\setcounter{equation}{0}
\section{Nuclear matrix elements at orders $q^3$ and $q^4$}
\label{MEq34}

In this appendix, we give the coordinate space representations corresponding
to the diagrams shown in Fig.~2. Whenever appropriate, we combine graphs.
Graph d vanishes because of isospin and e and f are proportional to
$v \cdot (q-k)$ which is zero at threshold to the order we are working.

\bigskip

\noindent a+k+o:
\begin{equation}
-\frac{ieg_A}{8\pi F_\pi^3}(1-2g_A^2)\, \int d^3r \, \phi^\ast (r) 
{\rm e}^{-i\frac{{\vec k} \cdot {\vec r}}{2}} \, O_1 \, 
\biggl\{ i \frac{{\vec k} \cdot {\vec r}}{r^3} - \frac{2}{r^2}
\frac{\partial}{\partial r} \biggr\} \, \phi(r) \,\, ,
\end{equation}

\noindent b+l+p:
\begin{eqnarray}
&& \frac{ieg_A}{8\pi F_\pi^3}(1-2g_A^2)\, \biggl[
\int d^3r \, \phi^\ast (r) {\rm e}^{i
\frac{{\vec k} \cdot {\vec r}}{2}} \biggl\{ \biggl[
-2 \, O_1  \, \frac{Y_1 (M_\pi r)}{r^2} + 2 \, O_2 \,  
\frac{Y_2 (M_\pi r)}{r^3} \biggr] 
\nonumber \\
&&+ \frac{i}{\pi} \int d^3r'
\bigl( \vec{\sigma\,}_1 + \vec{\sigma\,}_2 \bigr) \cdot {\vec r}\,' \,
\frac{Y_1 (M_\pi r')}{{r'}^2}\, \biggl[ \frac{{\vec \epsilon} \cdot
{\vec r}}{r}  \frac{-i}{|r+r'|^3} \frac{\partial}{\partial r} -
3 \frac{ \vec{\epsilon} \cdot ({\vec r} + {\vec r}\,' ) }{|r+r'|^5}
\nonumber \\
&& \qquad \quad \qquad \qquad 
\times\biggl(-i\frac{ ({\vec r} + {\vec r}\,' ) \cdot \vec{r} }{r}
\frac{\partial}{\partial r} + \frac{ ({\vec r} + {\vec r}\,' ) 
\cdot \vec{k} }{2} \biggr) \biggr] \,  {\rm e}^{i {\vec k} \cdot {\vec r}\,'}
\biggr\}  \, \phi(r) \, \biggr] \,\, ,
\end{eqnarray}
 
\noindent c+i+j:
\begin{eqnarray}
&&-i\frac{5eg_A}{8\pi F_\pi^3} \biggl[ -(1+\kappa_v) \,\int d^3r \, 
\phi^\ast (r)
\, O_3 \, \frac{Y_1 (M_\pi r)}{{r}^2} \,  {\rm e}^{i\frac{{\vec k} 
\cdot {\vec r}}{2}} \, \phi(r) \\
&&+\int d^3r \phi^\ast (r) \, {\rm e}^{-i\frac{{\vec k} \cdot {\vec r}}{2}}
\biggl\{ - O_1 \,\frac{Y_1 (M_\pi r)}{{r}^2} \, + \, O_2 \, 
\biggl( \frac{Y_2 (M_\pi r)}{r^3} -2 \frac{Y_1 (M_\pi r)}{r^3}
\frac{\partial}{\partial r} \biggr) \biggr\} \, \phi(r) \,\biggr] \,\, ,
\nonumber 
\end{eqnarray} 

\noindent g+h:
\begin{equation}
-\frac{ieg_A}{4\pi F_\pi^3} \int d^3r \, \phi^\ast (r) {\rm e}^{-i
\frac{{\vec k} \cdot {\vec r}}{2}}  
\biggl\{ -O_1 \, \frac{1}{r^3} + O_2 \, \biggl( \frac{3}{r^5} - \frac{2}{r^4}
\frac{\partial}{\partial r} \biggr) \biggr\} \, \phi(r) \,\, ,
\end{equation}

\noindent m+q:
\begin{eqnarray}
&& \frac{ieg_A^3}{4\pi F_\pi^3} \biggl\{
\int d^3r \, \phi^\ast (r) {\rm e}^{-i
\frac{{\vec k} \cdot {\vec r}}{2}} \biggl[
-O_1  \, \frac{Y_1 (M_\pi r)}{r^2} + O_2 \, \biggl(  \frac{Y_2 (M_\pi r)}{r^3} 
- 2\, \frac{Y_1 (M_\pi r)}{r^3} \frac{\partial}{\partial r} \biggr)\biggr]
 \, \phi(r)     \nonumber \\
&& \qquad \qquad \qquad \qquad
-  \int d^3r  \, \phi^\ast (r)  \biggl( i \,
O_4 \,\frac{Y_2 (M_\pi r)}{r^3} +  O_3 \,\frac{Y_1 (M_\pi r)}{r^2} \biggr)
\, {\rm e}^{i \frac{{\vec k} \cdot {\vec r}}{2}}
\, \phi(r) \, \biggr\} \,\,\, ,
\end{eqnarray}

\noindent n+r:
\begin{equation}
\frac{3eg_A^3  M_\pi }{8\pi^2 F_\pi^3}\,\int d^3r \, \phi^\ast (r) 
{\rm e}^{i \frac{{\vec k} \cdot {\vec r}}{2}} \,  O_4 \, \frac{1}{r^3} \,
\biggl[ - \frac{\partial K_0 (M_\pi r)}{\partial r} + r \,
\frac{\partial^2 K_0 (M_\pi r)}{\partial r^2} \, \biggr] 
\, \phi(r)  \,\,\, ,
\end{equation}

\noindent with
\begin{eqnarray}
Y_1 (M_\pi r) &=& \biggl(M_\pi + \frac{1}{r} \, \biggr) 
\, {\rm e}^{- M_\pi \, r} \, , \nonumber \\
Y_2 (M_\pi r) &=& \biggl(M_\pi^2 + \frac{3M_\pi}{r} + \frac{3}{r^2} \biggr) 
\, {\rm e}^{- M_\pi \, r} \, , \nonumber \\
K_0 (M_\pi r) &=& \int_0^\infty dz \frac{z \, \sin z}{(z^2 
+ M_\pi^2 r^2)^{3/2} } \, \, ,
\end{eqnarray}
and
\begin{eqnarray}
O_1 &=& \bigl( \vec{\sigma\,}_1 + \vec{\sigma\,}_2 \bigr) \cdot
\vec \epsilon  \,\, , \nonumber \\
O_2 &=& \bigl(\vec{\sigma\,}_1 + \vec{\sigma\,}_2 \bigr) \cdot {\vec r} 
\, {\vec \epsilon} \cdot {\vec r} \,\, , \nonumber \\
O_3 &=& \biggl( \vec{\sigma\,}_1 \cdot \vec{r} \, \vec{\epsilon} \cdot
({\vec k} \times \vec{\sigma\,}_2 \,) + (1 \leftrightarrow 2) \biggr) \,\, ,
\nonumber \\
O_4 &=& \biggl( \vec{\sigma\,}_1 \cdot \vec{r} \, \vec{\epsilon} \cdot
({\vec r} \times \vec{\sigma\,}_2 \,) + (1 \leftrightarrow 2) \biggr) \,\, ,
\end{eqnarray}
Performing the angular integrations, all the spin--dependence can be expressed
in terms of the spin vectors $\vec{\sigma}_{1,2}$ and the photon 
three--momentum $\vec k$  as given in Eq.(\ref{M12}).
The second operator in Eq.(\ref{M12}) stems in part from $O_2$,  
$O_3$ and $O_4$. The deuteron wave function $\phi( {\vec{r}} \,)$ is given by
\begin{equation}
\phi( {\vec{r}}\, ) = \frac{1}{\sqrt{4\pi}} \biggl( \frac{U(r)}{r} +
\frac{1}{\sqrt{8}} \, S_{12}({\hat{r}}) \, \frac{W(r)}{r} \biggr) \,\, ,
\end{equation}
in terms of the S-- and D--wave functions $U(r)$ and $W(r)$, respectively,
and 
\begin{equation}
S_{12} =3 ({\vec{\sigma}}_1 \cdot {\hat{r}}) ({\vec{\sigma}}_2
\cdot {\hat{r}}) - {\vec{\sigma}}_1 \cdot {\vec{\sigma}}_2
\end{equation}
is the tensor operator. The wave function is normalized to one, 
$\int_0^\infty dr \, (U^2 + W^2) = 1$. 

For completeness, we also give the corrected version of the $q^3$ 
contributions corresponding to graphs 1a and 1b. The angular
integrations are performed as outlined in ref.\cite{silas} using the
corrected relation
\begin{equation}
S_{12}^\ast \, (\vec J \cdot \vec \epsilon \, ) S_{12}
= 4 \, [ 3 (\vec \epsilon \cdot \hat r \,) \,(\vec J \cdot  \hat r \,)
- 2 \,(\vec J \cdot \vec \epsilon \, )] \,\,.
\end{equation}
This gives:

\noindent a:
\begin{eqnarray}
&&-\frac{ieg_A M_\pi m}{\pi F_\pi^3}\, \int \frac{dr}{ra^3} \,\biggl(
U^2 \, a^2 \, \sin a + \frac{1}{\sqrt{2}} \, U\, W (3 \sin a - 3a \cos a 
- a^2 \sin a ) \nonumber \\ && \qquad \qquad\qquad \qquad
 + \frac{1}{2} W^2 \,(3 \sin a - 3a \cos a -2a^2\sin a ) \biggr) \,\, ,
\end{eqnarray}

\noindent b:
\begin{eqnarray}
&&\frac{ieg_A M_\pi m}{\pi F_\pi^3}\, \biggl[ \, \int_0^1 dz
 \int dr \,\frac{{\rm e}^{-m'r} }{rb^3} \,\biggl(
U^2 \, b^2 \, \sin b + \frac{1}{\sqrt{2}} \, U\, W (3 \sin b - 3b \cos b
- b^2 \sin b ) \nonumber \\ && \qquad \qquad\qquad \qquad
 + \frac{1}{2}W^2 \,(3 \sin b - 3b \cos b -2b^2\sin b ) \biggr) 
\nonumber \\ && -\int_0^1 dz \int\, dr \, Y_1(m'r) \frac{1}{b^3} \,
(\sin b - b \cos b) \biggl(U + \frac{1}{\sqrt{2}} W \biggr)^2
\\
&& -\frac{3 |\vec{k}\, |}{\sqrt{2}} \, \int_0^1 dz \, (z-1) \,
\int dr \,\frac{{\rm e}^{-m'r} }{b^4} \biggl( U\,W + 
\frac{1}{\sqrt{2}} W^2 \biggr) \,((3-b^2)\sin b - 3b \cos b) \,
\biggr]\,\, , \nonumber 
\end{eqnarray}
with
\begin{equation}
a = \frac{kr}{2} \, , \quad b = kr \biggl( z -\frac{1}{2} \biggr) \, ,
\quad m' = M_\pi \, \sqrt{z \, (2-z)} \,\, \, .
\end{equation}

\vfill\eject

\setcounter{equation}{0}
\section{Time orderings in heavy fermion formalism}
\label{time}

In this appendix we establish a simple method for extracting
irreducible time ordered graphs in baryon chiral perturbation theory.
Consider the reducible relativistic Feynman graph shown in Fig.~4.
Since there are $3$ interaction vertices we expect $3!$ time
orderings. The time ordered decomposition of Fig.~4 is displayed in
Fig.~5. There are several important points to note:

\vspace{0.12in}
\noindent$\bullet$
Graphs $(4)$, $(5)$ and $(6)$ are $1/m$ corrections and so
the sum of $(1)$, $(2)$ and $(3)$ correspond to the Feynman graph of
Fig.~4 evaluated in the heavy fermion formalism (HFF).
\vspace{0.05in}

\noindent$\bullet$
Graphs $(2)$ and $(3)$ are reducible and therefore
graph $(1)$ is the irreducible subgraph that we are after.
\vspace{0.12in}

The propagator structure of Fig.~4 is

\begin{eqnarray}
&&
\frac{i}{{k^2}-{M_\pi^2}}\,\frac{i({\not \! p}+m)}{{p^2}-{m^2}}=
\frac{i}{2\omega}
\biggl(\frac{1}{{k_0}-\omega}+\frac{-1}{{k_0}+\omega}\biggr)\nonumber \\
&&\qquad\qquad\qquad\qquad\,
\times\frac{i}{2{E_n}}
\biggl(
\frac{{\gamma_0}{E_n}-\vec\gamma\cdot{\vec p}+m}{{p_0}-{E_n}}+
\frac{{\gamma_0}{E_n}+\vec\gamma\cdot{\vec p}-m}{{p_0}+{E_n}}
\biggr)
\end{eqnarray}
where ${E_n}=\sqrt{{{\vec p}^2}+{m^2}}$
and $\omega=\sqrt{{{\vec k}^2}+{M_\pi^2}}$.
We are interested in positive frequency pions and nucleons
and so

\begin{equation}
(1)+(2)\propto \frac{-1}{4\omega{E_n}}
\biggl(\frac{1}{{k_0}-\omega}\,
\frac{{\gamma_0}{E_n}-\vec\gamma\cdot{\vec p}+m}{{p_0}-{E_n}}
\biggr).
\end{equation}
This expression can be partial fractioned to give

\begin{equation}
\frac{-{\gamma_0}{E_n}+\vec\gamma\cdot{\vec p}-m}{4\omega{E_n}}\,
\frac{1}{{k_0}+{p_0}-{E_n}-\omega}
\biggl(\frac{1}{{k_0}-\omega}+\frac{1}{{p_0}-{E_n}}\biggr).\nonumber \\
\end{equation}
In order to show that the two pieces inside the parentheses are in
correspondence with $(1)$ and $(2)$, we evaluate $(1)$ and $(2)$ using
old-fashioned time ordered perturbation theory. Here we attach a
photon and a pion line to our generic interaction vertex and focus on
the distinct time slices, labelled $(i)$ and $(ii)$ (see Fig.~6). The
rules are simple: each distinct time slice ---with respect to the
interaction vertices--- corresponds to a single energy propagator. We
find

\begin{eqnarray}
(1)&&\propto\,
\frac{1}{[({E_1}+{E_2}+{E_\gamma})-({E_1'}+\omega+{E_n}+{E_\pi})]}
\qquad\qquad (i)
\nonumber \\
&&\times\frac{1}{[({E_1}+{E_2}+{E_\gamma})-({E_1'}+\omega+{E_2}+{E_\gamma})]}
\qquad\qquad\, (ii)
\end{eqnarray}

\begin{eqnarray}
(2)&&\propto\,
\frac{1}{[({E_1}+{E_2}+{E_\gamma})-({E_1'}+\omega+{E_n}+{E_\pi})]}
\qquad\qquad (i)
\nonumber \\
&&\times\frac{1}{[({E_1}+{E_2}+{E_\gamma})-({E_1}+{E_n}+{E_\pi})]}
\qquad\qquad\qquad (ii).
\end{eqnarray}
In relativistic notation we have
${k_0}={E_1}-{E_1'}$ and ${p_0}={E_2}-{E_\gamma}-{E_\pi}$.
(Note that energy is {\it not} conserved at the vertices in the
time ordered formalism.) It then follows that

\begin{eqnarray}
&&(1)\propto \frac{1}{{k_0}+{p_0}-{E_n}-\omega}\times
\frac{1}{{k_0}-\omega}\nonumber \\
&&(2)\propto \frac{1}{{k_0}+{p_0}-{E_n}-\omega}\times
\frac{1}{{p_0}-{E_n}}.
\end{eqnarray}

We are now in position to establish a rule for extracting the
irreducible time ordered subgraph, $(1)$, in HFF. The propagator structure of
Fig.~4 in HFF is

\begin{equation}
\frac{i}{{k^2}-{M_\pi^2}}\,\frac{i}{v\cdot k}
\end{equation}
where $p=mv+k$. In the static limit ${v\cdot k}\to{k_0}$=$0$, which renders
(D.7) singular. In this limit ${E_n}=m$ and ${p_0}=m-{k_0}$. 
Therefore, decomposing the propagators gives

\begin{eqnarray}
\frac{i}{{k^2}-{M_\pi^2}}\,\frac{i}{k_0}&&=
\frac{i}{2\omega}\biggl(\frac{1}{{k_0}-\omega}+\frac{-1}{{k_0}+\omega}\biggr)
\times \frac{-i}{{p_0}-{E_n}}\nonumber \\
&&=\frac{1}{2\omega}\,\frac{1}{({k_0}-\omega )({p_0}-{E_n})}
+(-)\, {\it T.O.}\nonumber \\
&&=\frac{1}{2\omega}\,\frac{1}{{k_0}+{p_0}-{E_n}-\omega}
\biggl(\frac{1}{{k_0}-\omega}+\frac{1}{{p_0}-{E_n}}\biggr)
+(-)\, {\it T.O.}\nonumber \\
&&=\frac{1}{2\omega}\,\frac{1}{({k_0}+{p_0}-{E_n}-\omega)}
\frac{1}{({k_0}-\omega )}
+{\it reducible}\, (+)\, {\it T.O.}
+(-)\, {\it T.O.}
\nonumber \\
\end{eqnarray}
where $(\pm)$ refers to the frequency of the time ordering and
(D.6) was used in the last line.
The first term is graph $(1)$ of Fig.~5, the {\it irreducible}
$(+)$ {\it T.O.} piece. It is, of course, well behaved in
the static limit. In the limit ${k_0}\to 0$ we have

\begin{equation}
{\it irreducible}\,(+)\, {\it T.O.}
=\frac{1}{2{\omega}^3}=\frac{1}{2{({{\vec k}^2}+{M_\pi^2})}^{{3}/{2}}}
\end{equation}
and so the irreducible time ordering in a HFF Feynman graph
of type Fig.~4 can be extracted by making the following replacement
in the HFF propagators:

\begin{equation}
\frac{i}{{k^2}-{M_\pi^2}}\,\frac{i}{v\cdot k}\Longrightarrow
\frac{1}{2{({{\vec k}^2}+{M_\pi^2})}^{{3}/{2}}}.
\end{equation}


\bigskip \bigskip 


\section*{Figures}

\begin{figure}[h]
   \vspace{0.5cm}
   \epsfysize=5cm
   \centerline{\epsffile{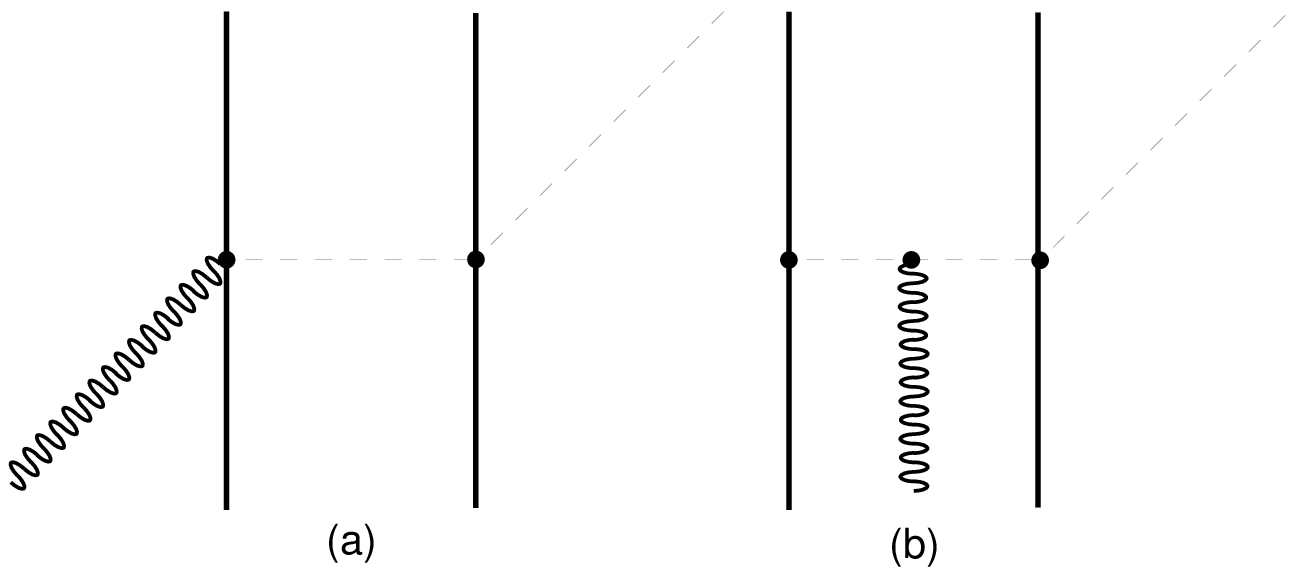}}
   \centerline{\parbox{11cm}{\caption{\label{fig1}
Three--body interactions which contribute to neutral pion photoproduction
at threshold to order $q^3$ (in the Coulomb gauge).
  }}}
\end{figure}

\begin{figure}[t]
   \vspace{0.5cm}
   \epsfysize=14cm
   \centerline{\epsffile{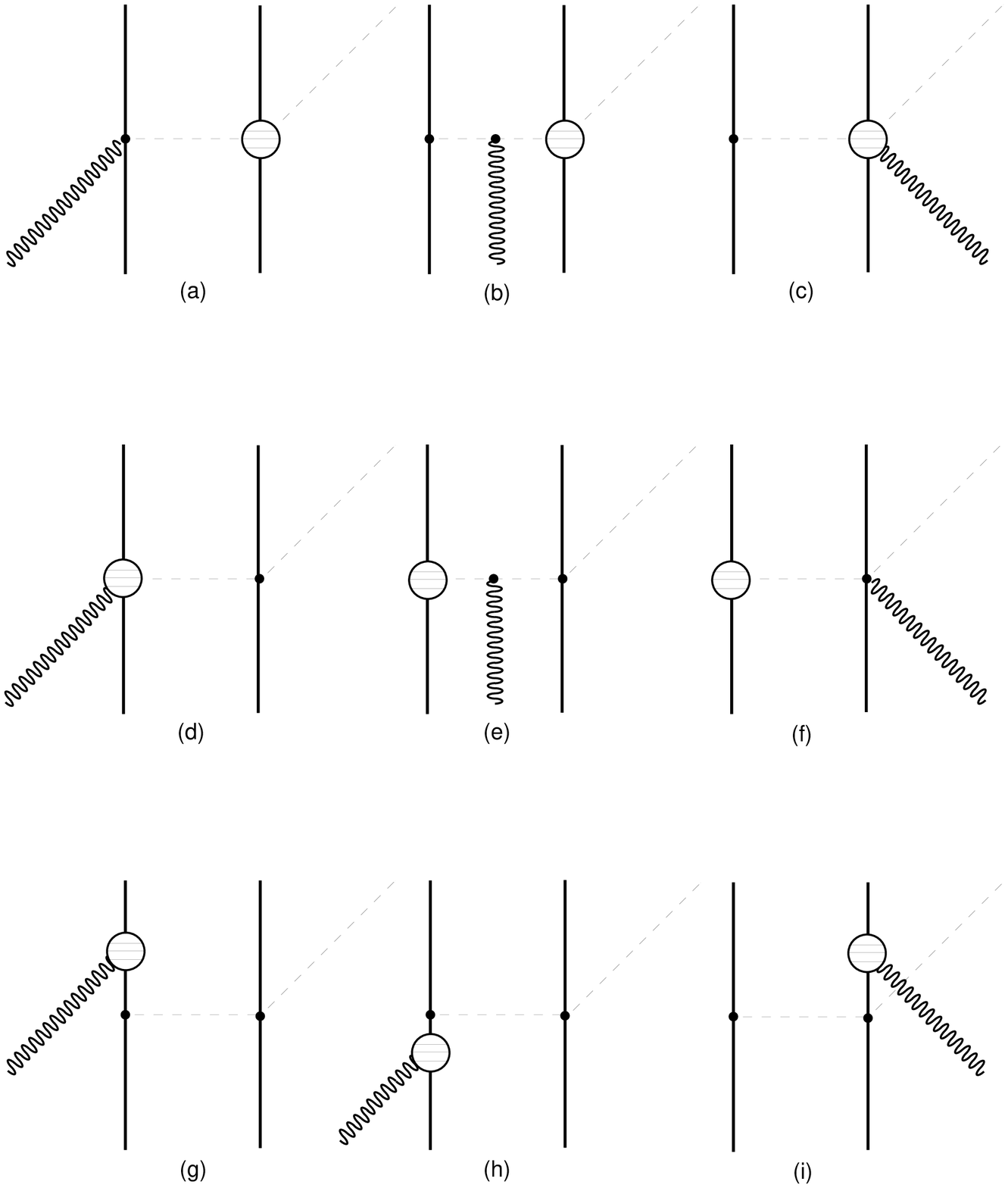}}
\end{figure}

\begin{figure}[t]
   \vspace{0.5cm}
   \epsfysize=14cm
   \centerline{\epsffile{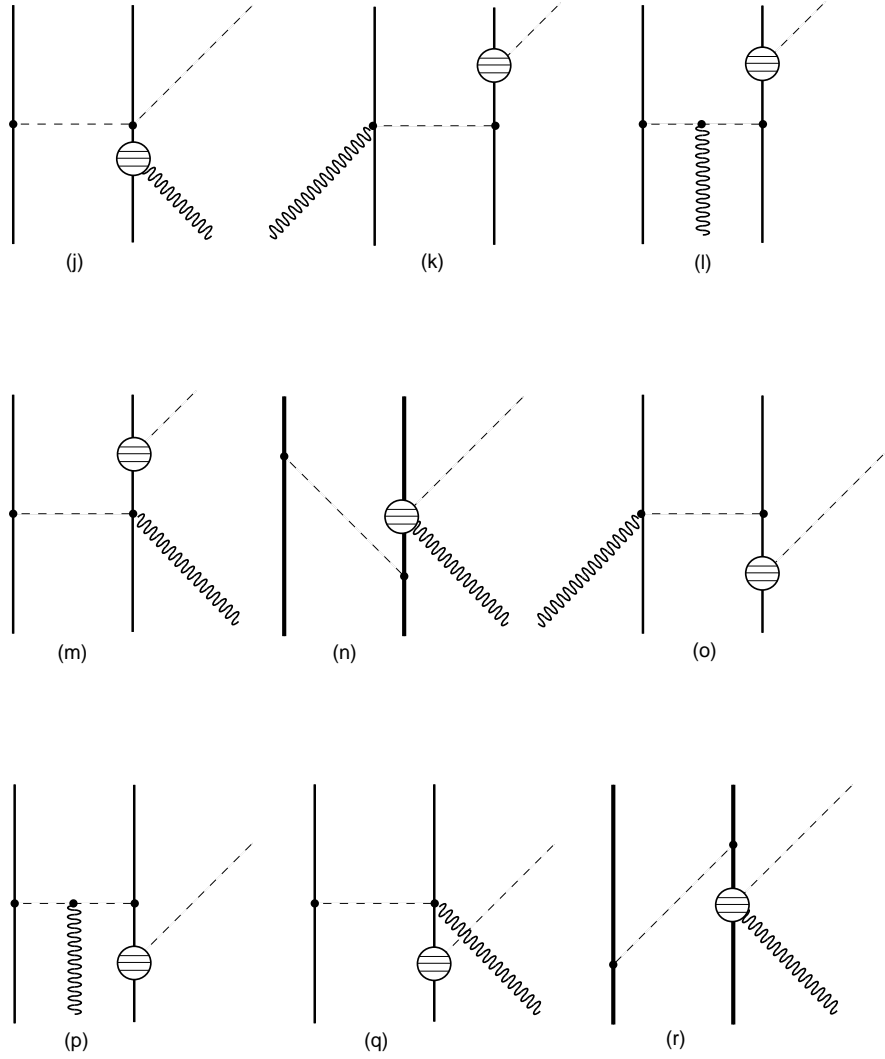}}
   \centerline{\parbox{11cm}{\caption{\label{fig2a}
Graphs contributing at order $q^4$ to neutral pion photoproduction.
The hatched circles denote an insertion from ${\cal L}_{\pi N}^{(2)}$.
The time--ordered graphs are distinguished by bold nucleon lines.
  }}}
\end{figure}

\begin{figure}[t]
   \vspace{0.5cm}
   \epsfysize=5cm
   \centerline{\epsffile{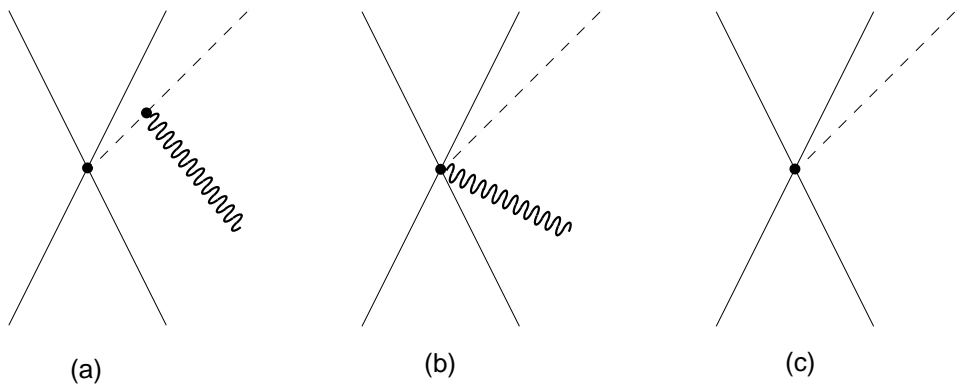}}
   \centerline{\parbox{11cm}{\caption{\label{fig3}
Four--fermion terms contributing to (charged) pion
photoproduction for $\nu =0$ (graphs a and b). Graph b
is generated from graph c by minimal substitution.
  }}}
\end{figure}

\begin{figure}[b]
   \vspace{0.5cm}
   \epsfysize=6cm
   \centerline{\epsffile{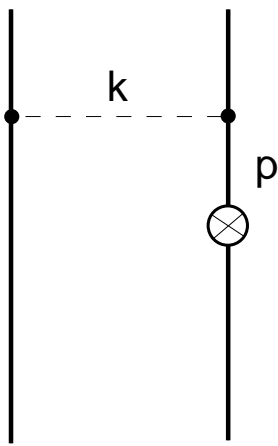}}
   \centerline{\parbox{11cm}{\caption{\label{fig4}
A two-nucleon reducible relativistic Feynman graph. The cross denotes
a generic interaction vertex.
  }}}
\end{figure}

\begin{figure}[t]
   \vspace{0.5cm}
   \epsfysize=10cm
   \centerline{\epsffile{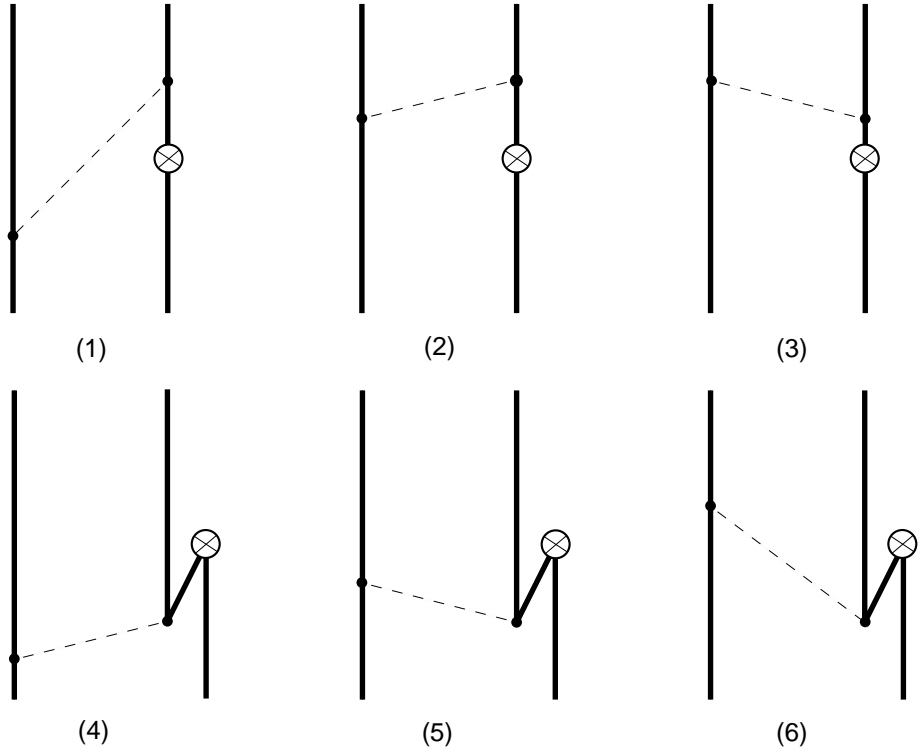}}
   \centerline{\parbox{11cm}{\caption{\label{fig5}
Decomposition of Fig.~4 into distinct time orderings.
Graphs $(4)$, $(5)$ and $(6)$ are $1/m$ corrections and so
the sum of $(1)$, $(2)$ and $(3)$ is the Feynman graph in HFF.
  }}}
\end{figure}

\begin{figure}[b]
   \vspace{0.5cm}
   \epsfysize=6cm
   \centerline{\epsffile{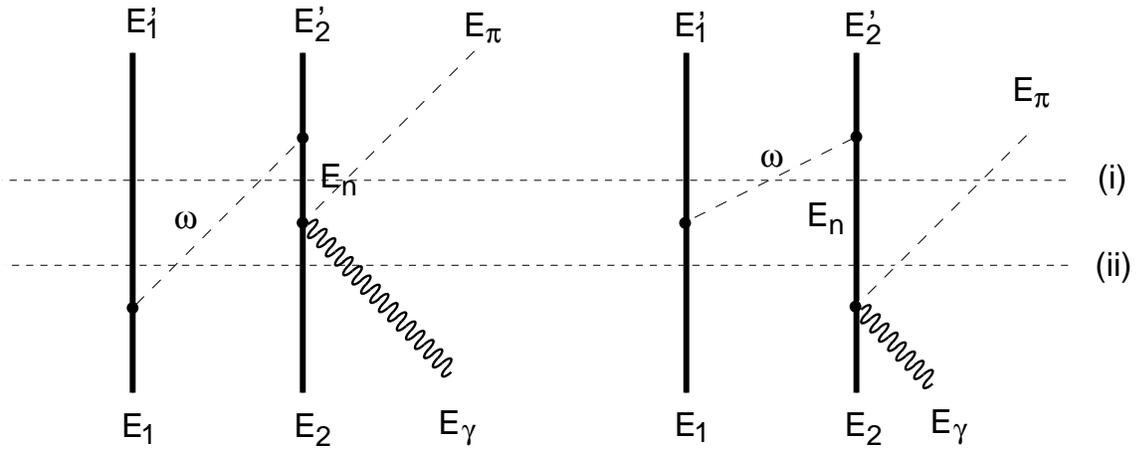}}
   \centerline{\parbox{11cm}{\caption{\label{fig6}
Time slices for graphs (1) and (2) of Fig.~5 with pion and
photon attached. All energies flow from bottom to top.
  }}}
\end{figure}

\end{document}